\documentclass[preprint,12pt]{elsarticle}

\usepackage{amsmath}
\usepackage{amssymb}
\usepackage{color}
\journal{Physica A}

\begin{document}

\begin{frontmatter}

\title{Dynamic-sensitive cooperation in the presence of multiple strategy updating rules}

\author[mfa]{Attila Szolnoki}
\author[nyf]{Zsuzsa Danku}
\address[mfa]{Institute of Technical Physics and Materials Science, Centre for Energy Research, Hungarian Academy of Sciences, P.O. Box 49, H-1525 Budapest, Hungary}
\address[nyf]{Institute of Mathematics and Informatics, H-4401 Ny\'{\i}regyh\'aza, Hungary}

\begin{abstract}
The importance of microscopic details on cooperation level is an intensively studied aspect of evolutionary game theory. Interestingly, these details become crucial on heterogeneous populations where individuals may possess diverse traits. By introducing a coevolutionary model in which not only strategies but also individual dynamical features may evolve we revealed that the formerly established conclusion is not necessarily true when different updating rules are on stage. In particular, we apply two strategy updating rules, imitation and Death-Birth rule, which allow local selection in a spatial system. Our observation highlights that the microscopic feature of dynamics, like the level of learning activity, could be a fundamental factor even if all players share the same trait uniformly.
\end{abstract}

\begin{keyword}
\texttt cooperation\sep\ social dilemma\sep\ imitation\sep\ death-birth rule
\end{keyword}

\end{frontmatter}

\section{Introduction}

Cooperation among self-interested actors is a widespread phenomenon that bridges several otherwise disjunct disciplines \cite{axelrod_84,ostrom_90,sigmund_10,bowles_11,nowak_11}. This seemingly paradoxical behavior, where individual and collective interests are in conflict, provides a major challenge for our century \cite{pennisi_s05}. Not surprisingly, an armada of scientists with different backgrounds are trying to identify decisive factors which may explain the evolutionary success of altruistic choice \cite{szabo_pr07,archetti_jtb12,fu_pre08,chen_xj_jtb13,perc_bs10, chen_xj_pre08b,takesue_epl17,wu_t_pcbi17,perc_pr17}. Albeit a clear and intuitive taxonomy of potential cooperation supporting mechanisms was already given by Martin Nowak \cite{nowak_s06}, but this framework can mostly be considered as an inspiring starting point for subsequent research efforts.

One of the main research paths reveals the possible diverse consequences of different strategy updating rules on the evolution of competing strategies \cite{ohtsuki_jtb06,roca_plr09,zukewich_pone13,hindersin_pcbi15}. When imitation is used, which is the most frequently applied strategy updating rule \cite{szabo_pre98}, it turned out that the accompanying individual features, such as strategy learning or teaching capacity, 
could be a decisive factor if we assume a heterogeneous population where actors may differ from each other \cite{szolnoki_epl07}. Even if we assume diverse actors the consequences of varying learning or varying teaching capacities are largely different. While diverse learning activity has no particular role on cooperation level the possibility of unequal teaching activity allows network reciprocity to be augmented in a similar way that was observed previously for largely heterogeneous interaction graphs \cite{santos_prl05}. More precisely, a player having large strategy teaching capacity is able to enforce her strategy to her local neighborhood \cite{szolnoki_epjb08}. In this way locally coordinated homogeneous spots emerge which reveals the advantage of mutual cooperation. It is crucial to stress, however, that the benefit of individual strategy teaching capacity is only visible if players are heterogeneous, but disappears if players are uniform and bear identical dynamical features. In the latter case the final evolutionary outcome is independent of the proper value of the uniformly applied teaching or learning activity.

In this work we explore weather the individual dynamical features of actors, like strategy learning or strategy teaching capacities, play any role on the evolution of cooperation when there are different ways to update their strategies. For this purpose we assume that both imitation and Death-Birth (DB) updating rules are available and we also suppose that the mentioned dynamical features of players may change individually \cite{ohtsuki_jtb06}. In particular, we assume that players may have lower or higher strategy learning capacities which may change during the evolutionary process. In an alternative setup we assume distinct strategy teaching activities and clarify whether they affect the evolutionary outcome when both imitation and Death-Birth rules are present. We stress that both mentioned updating protocols assume only information about local neighborhood hence their alternative use does not cause fundamental differences. The latter would not be hold for those updating rules which use global information about the entire population. For example, when Birth-Death (BD) updating is used a player is chosen for reproduction from the entire population proportional to fitness. In a similar way, global information, the average level of fitness, is necessary when replicator dynamics is applied. To avoid incomparable features of updating protocols we only use imitation and DB rule.

We note that the simultaneous use of different strategy updating rules can be introduced in two basically distinct ways. According to the first scenario, which is conceptually similar to an annealed randomness, a player may use either imitation or can be the subject to a Death-Birth process, but these updating protocols are used with a specific probability. In the other case, which resembles quenched randomness, a player uses one of the mentioned updating rules exclusively, but the fraction of those players who belongs to a specific set is well-defined. We note that considering an adaptive population where players playing a skill game was reported in \cite{javarone_jsm15}.

In the following we will show that the way how we mix the updating rules can be a decisive factor on the resulting cooperation level. Interestingly, the level of strategy learning capacity is more important factor than the strength of teaching capacity, which is against our previous experience when only a single strategy updating protocol was present \cite{szolnoki_epl07}.

The remainder of this paper is organized as follows. In the next Section we describe the model and survey the possible versions of the proposed evolutionary games. Section~\ref{result} is devoted to the presentation of our observations. Finally, we summarize the main conclusions and discuss their potential implications in Section~\ref{disc}.

\section{Model and Method}
\label{def}

We consider a population of individuals who play the so-called weak prisoner's dilemma game on a graph \cite{nowak_n92b}. In this simplified version, which still captures the essence of a social dilemma, we only have a single parameter $T$ that characterizes the strength of the dilemma. Initially each player on site $i$ is designated as either a cooperator ($C$) or a defector ($D$) with equal probability. While mutual
cooperation yields the reward $R = 1$ to both cooperator players, mutual defection results in zero payoff to the partners. The same zero payoff goes to a cooperator who interacts with a defector, while the latter collects the temptation $T>1$ value, which establishes that defection is the preferred individual choice.

For simplicity we present our results obtained on a $L\times L$ square lattice with periodic boundary conditions, but we stress that qualitatively similar behavior can be found by using other types of topologies including regular and heterogeneous random graphs \cite{watts_dj_n98,szabo_jpa04}. According to the applied interaction topology, when calculating the payoff of a player then we accumulate the payoff values obtained from the pair interactions with all neighbors. 

In every Monte Carlo step in average all players have a chance to update their strategies. During the strategy updating protocol we use Death-Birth process with probability $q$ and imitation rule with probability $1-q$. In the former case a randomly chosen individual is removed and her neighbors compete for the empty site proportional to their fitness. In the alternative case, which happens with probability $1-q$, the imitation rule is considered. Accordingly, the randomly selected player $i$, having strategy $s_i$, imitates the $s_j$ strategy of a neighboring player $j$ with a probability $\Gamma(s_j \to s_i)= 1/ \{1+\exp[(\Pi_i-\Pi_j) /K]\}$. Here $\Pi$ denotes the accumulated payoffs of both players while parameter $K$ quantifies the uncertainty of strategy adoptions \cite{szabo_pre98}. To gain results comparable to previous studies we apply $K=0.1$, but our observations remain intact for wide range of noise interval.

It is crucial to note that the above mentioned strategy updating rules use local selection, which is a fundamental feature when spatially structured population is considered. In other words, we don't need global information, {\it i.e.} to know the accurate states of all players in the whole population, when a microscopic strategy update is executed.

The above specified mixture of strategy updating rules resembles annealed randomness in statistical physics \cite{landau_00}. We also introduce an alternative way how to apply Death-Birth and imitation rules simultaneously. In the latter case, which is conceptually similar to quenched randomness in solid-state physics, a specific player always uses one of these rules. But the fraction of those sites where Death-Birth is applied is $q$, while the remaining $1-q$ portion of the population use imitation to update their states. 

Since our principal interest is to clarify whether an individual dynamical feature influences the final outcome we also introduce a certain trait which determines the success of the imitation process locally. For example, we can assume that players have different abilities to learn from their neighbors, which can be described by a $0 < w_l \le 1$ parameter. Consequently, when a player $i$ imitates a player $j$ then we assume a modified imitation probability that is $\Gamma^\prime (s_j \to s_i)= w_l(i)/ \{1+\exp[(\Pi_i-\Pi_j) /K]\}$. For simplicity, we assume that two different values, $w_l < 1$ and $w_l = 1$, are available in the initial state and these individual learning capacities are also adopted during the imitation process. The key questions are whether different individual traits coexist in the stationary state and how the specific value of $w_l$ influences the cooperation level. We note that the value of $w_l$ has no any significance in the classical model where all players use the same value and individual state can be varied via imitation only.

Evidently, individual strategy teaching capacity can also be introduced. In this case  when a player $i$ imitates a player $j$ then we assume a modified imitation probability that is $\Gamma^\prime (s_j \to s_i)= w_t(j)/ \{1+\exp[(\Pi_i-\Pi_j) /K]\}$. Again, for simplicity, we assume that two different values, $w_t < 1$ and $w_t = 1$, are available in the initial state and these individual teaching capacities are also adopted during the imitation process.

To summarize the model definition, practically we study four fundamentally different setups. In the first case we assume cooperator and defector players who may have different learning capacities and the microscopic updating process is executed via an annealed mixture of Death-Birth and imitation rules. In the second case we use quenched mixture of updating rules for the same players. Thirdly, we assume players who have different teaching activities with annealed mixture of updating rules. Finally, the players with heterogeneous teaching capacities are updated by a quenched mixture of microscopic protocols.

\section{Results}
\label{result}

\begin{figure}[h!]
\centering
\includegraphics[width=6.5cm]{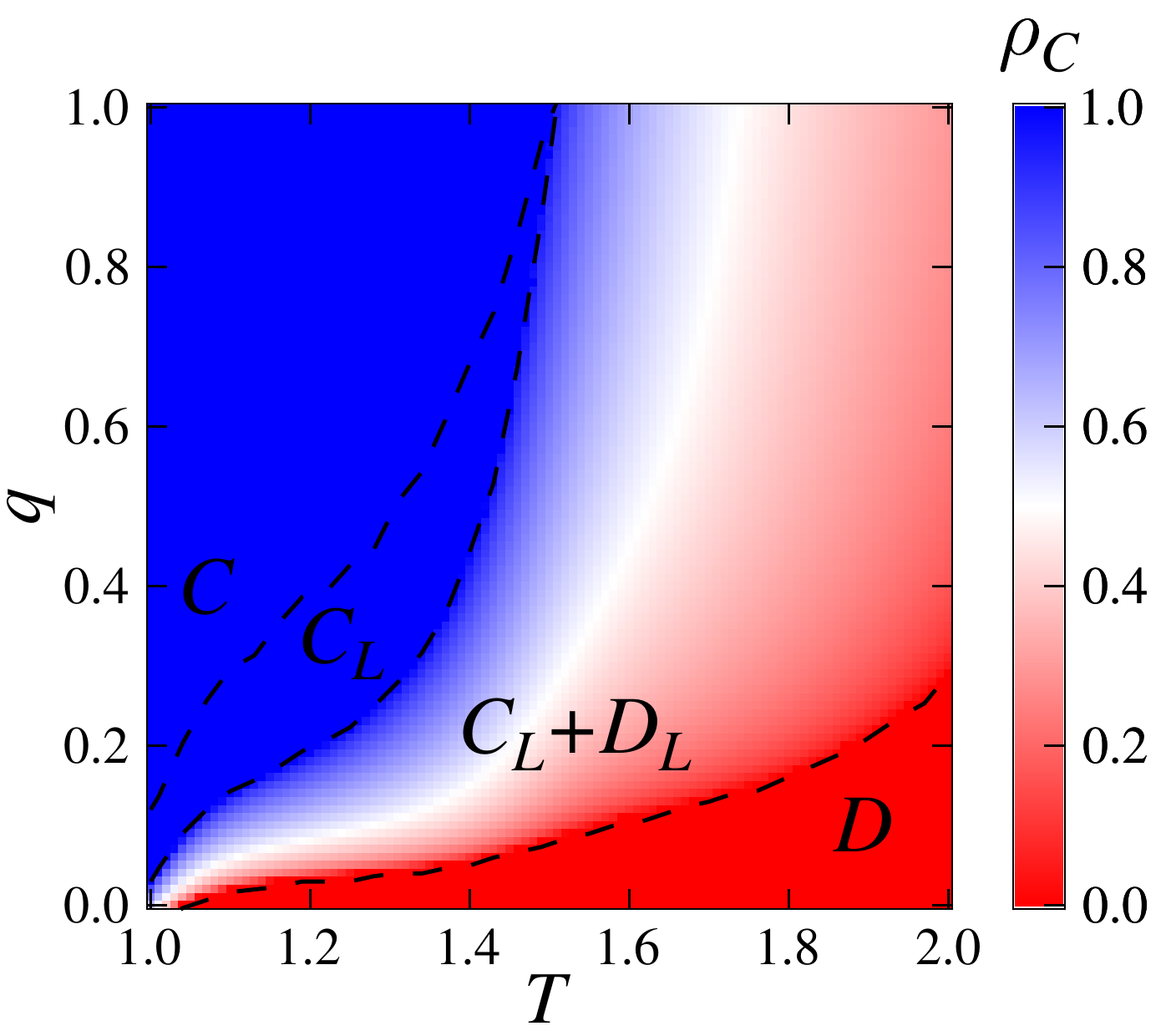}\includegraphics[width=6.5cm]{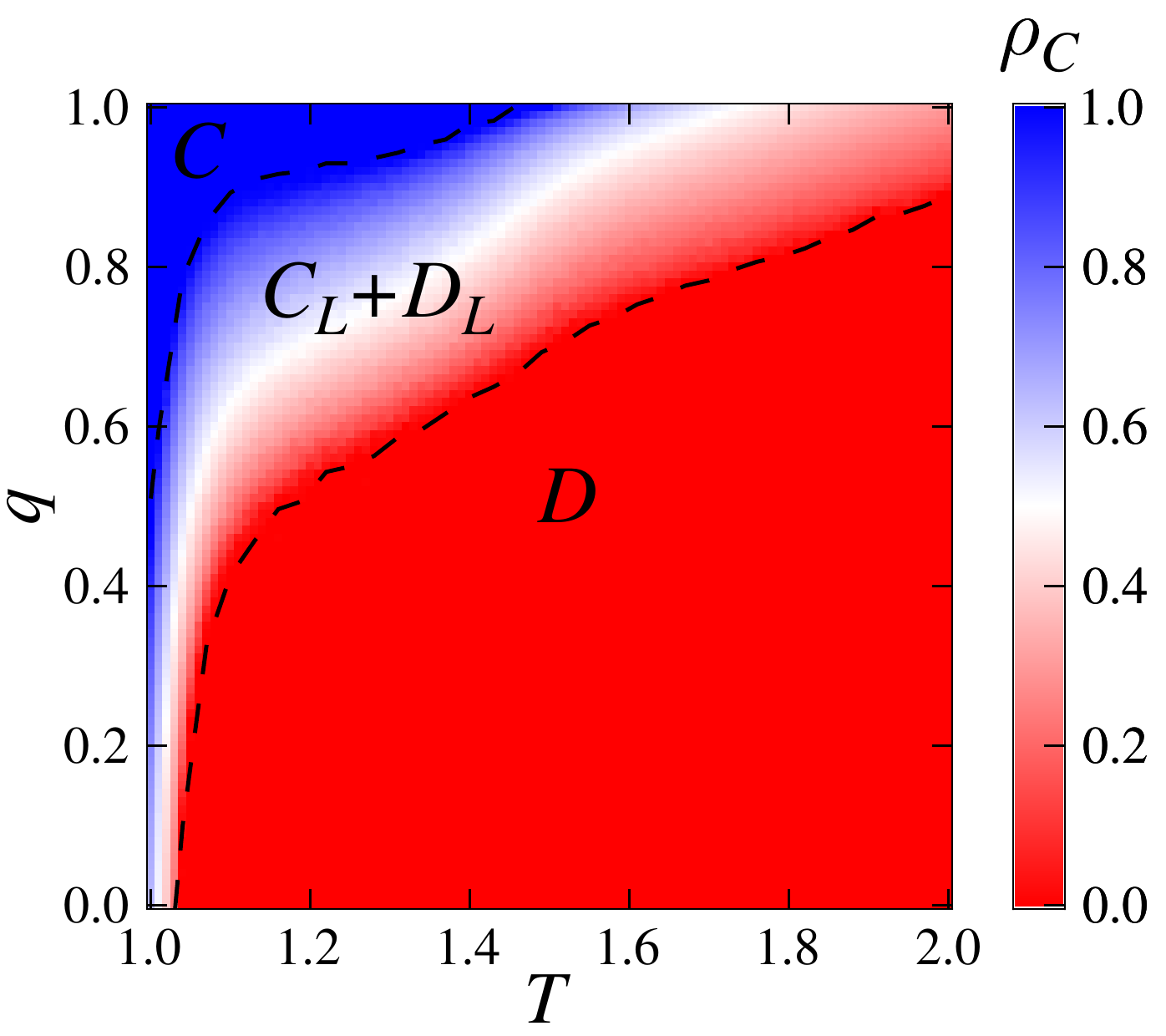}\\
\caption{Phase diagram of the coevolutionary model with heterogeneous learning activity obtained for annealed (left) and quenched (right) mixture of updating rules when $w_l=0.1$ is used beside the usual $w_l=1$ value. The applied microscopic dynamics is explained in the main text. $D$ denotes the phase where cooperators with different learning capacities die out first and only defectors remain. Alternatively, $C$ denotes the parameter region where defectors go extinct leaving back cooperators with different learning capacities. Interestingly, $C_L$ marks also a full cooperator phase, but here defectors and cooperators with higher learning capacity die out simultaneously. In the intermediate phase, where cooperators and defectors coexist, only players with low learning activity survive. The results were obtained up to $L=2000$ system size and remain robust in the large size limit.}\label{phd_learn}
\end{figure}

We start with the presentation of key findings for the first two cases. These results are summarized in the phase diagrams plotted in Fig.~\ref{phd_learn}. The control parameters are the $T$ temptation value and the $q$ probability that characterizes the weight of Death-Birth update in the microscopic dynamics. As we noted, here four different microscopic states compete for space during the coevolutionary process. They are cooperators with low ($C_L$) and high ($C_H$) learning capacities and defectors with low ($D_L$) and high ($D_H$) imitation skill. 

It is a mutual feature of both phase diagrams that players with high learning activity cannot survive in the stationary states. The only exception is at high $q$ -- low $T$ values where defectors die out first leaving behind cooperator players with different learning capacities. We marked this uniform-strategy state by ``$C$" where learning capacity of players becomes irrelevant. The system arrives to a similar uniform-strategy state at low $q$ -- high $T$ parameter values where cooperators die out first and defectors with different learning skills remained alive (this state is marked by ``$D$" on the diagrams).

When the updating rules are mixed in an annealed way, shown in the left panel of Fig.~\ref{phd_learn}, then the positive effect of the Death-Birth updating rule can be observed already at small $q$ values. Here the full cooperator state can be easily reached even at relatively high $T$ values. Interestingly, $C_L$ marks a phase where players with high learning capacities die out first that is followed by $D_L$ players and finally $C_L$ players prevail exclusively. If we increase the temptation at a fixed $q$ value then the system terminates into a mixed state where $C_L$ and $D_L$ players coexist. Nevertheless, the most striking feature of the diagram is that the influence of Death-Birth process emerges very early even at small $q$ values if we initially allow players to be present with different learning capacities.

The above described effect is completely missing if we mix the updating rules in a quenched way. This case is summarized in the right panel of Fig.~\ref{phd_learn}. Here the positive consequence of the presence of Death-Birth rule emerges only at high $q$ values where those players who follow this protocol percolate, hence they can support each other mutually.

\begin{figure}
\centering
\includegraphics[width=13.5cm]{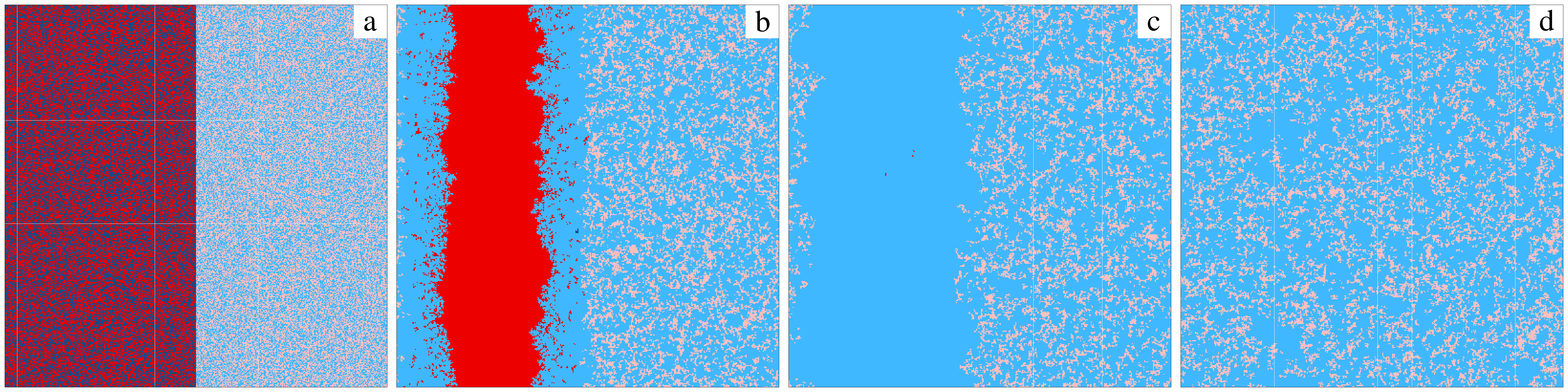}\\
\caption{Coevolution of strategies and learning capacities at $T=1.3$ and $q=0.2$. Panel~(a) shows the starting state where players with high learning capacities ($w_l=1$) are distributed in the left side of the space. Here $D_H$ defector players (dark red) and $C_H$ cooperators (dark blue) are present. The right side of the initial panel is filled with players with low learning capacities ($w_l=0.1$). They are $D_L$ defectors (light red) and $C_L$ cooperators (light blue). When evolution is launched, shown in panel~(b), $D_H$ players invade $C_H$ players. In parallel, $C_L$ players beat $D_H$ players, which will result in the extinction of both $C_H$ and $D_H$ states. In the final stationary state, shown in panel~(d), $C_L$ and $D_L$ players coexist despite of the high temptation value. Linear system size is $L=400$.}\label{learn_a_t}
\end{figure}

To gain a deeper insight about the microscopic mechanisms which govern the pattern formation we present a characteristic spatial evolution of the four competing states for both cases. In particular, we show the case of annealed mixture of updating rules, where  some representative snapshots are plotted in Fig.~\ref{learn_a_t}. The first panel of Fig.~\ref{learn_a_t} shows a prepared initial state where we separated players with different learning capacities in different sides of the available space. More precisely, players having high learning capacities ($w_l=1$) are separated on the left side. They are $D_H$ defectors (dark red) and $C_H$ cooperator players (dark blue). Initially, on the right side are those who have low, $w_l=0.1$, learning capacities. Here $D_L$ defectors are denoted by light red, while $C_L$ cooperators by light blue. When the evolution is launched the two subsystems evolve very differently. While $D_H$ players invade $C_H$ players efficiently, resulting a homogeneous dark red domain, $D_L$ and $C_L$ players coexist on the right side. Simultaneously, light blue $C_L$ players start invading $D_H$ opponents as it is shown in panel~(b). It will result in the extinction of $D_H$ state, shown in panel~(c), but the homogeneous $C_L$ domain is unstable due to the high $T$ and low $q$ value. The final stable state is shown in panel~(d) where low learning activity defectors and cooperators coexist. (To monitor the whole evolution we provided an animation that can be seen in \cite{learn}.) 
This example illustrates nicely that the individual dynamical feature, {\it viz.} learning capacity, can play a decisive role on the stationary state even if it is uniform for all players. The important condition is the presence of multiple updating rules: while cooperators cannot survive when players possess high learning capacities, they coexist with defectors if $w_l$ is low, no matter we applied a significantly strong temptation value.

The above described mechanism cannot work for quenched mixture of updating rules because the positive consequence of Death-Birth rule is inefficient when the sites which use this protocol are rare. If they cannot percolate then they are surrounded by sites where only imitation is used. The latter updating rule provides a significantly modest cooperation level and cooperators die out at every reasonable $T$ value. It simply means that only defectors will compete for the empty place where Death-Birth rule is used, hence the total failure of cooperation is inevitable. This situation can only change when Death-Birth places are dense enough to percolate. Above the percolation threshold \cite{landau_00} these sites become neighboring and their neighborhood should not necessarily follow the evolutionary trajectory dictated by pure imitation dynamics. In agreement with this argument, the right panel of Fig.~\ref{phd_learn} illustrates nicely that cooperation can only maintain in the high $q$ regime.

\begin{figure}
\centering
\includegraphics[width=13.5cm]{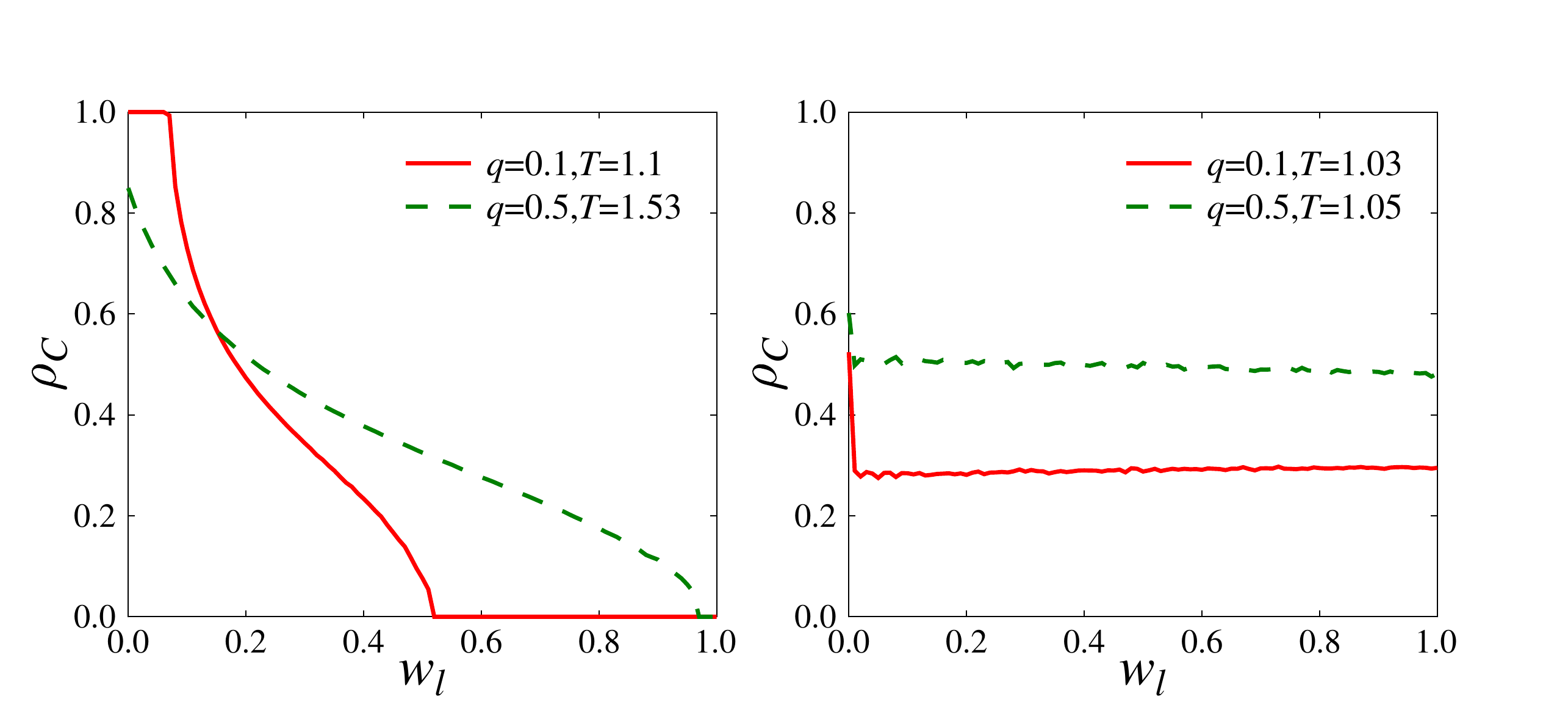}\\
\caption{Cooperation level in dependence of the suppressed learning capacity for annealed (left) and quenched (right) mixture of updating rules. The applied parameter values are shown in the legend. While suppressed learning capacity has a decisive role on cooperation level for the annealed case, it is irrelevant for the other case when players use permanently a certain type of updating rules.}\label{learn_compare}
\end{figure}

Put differently, the quenched mixture of updating rules does not provide any synergistic effect and practically we have two ``subsystems" where either imitation or Death-Birth-type updating rule is functioning. (We should not forget that the latter is functioning properly only at high $q$ values.) Since we practically have ``subsystems" using a single-updating rule, therefore the individual dynamical feature becomes unimportant again. As we already mentioned in the introduction, in the traditional model, where only imitation is used, the actual value of $w_l$ plays no any role if all actors possess the same trait \cite{szolnoki_epl07}. According to this argument we should obtain the same evolutionary outcome independently of the applied $w_l$ value if quenched mixture of updating rules is applied. This conjecture is nicely confirmed in Fig.~\ref{learn_compare} where we plotted the cooperation level in dependence of $w_l$ lower teaching activity both for annealed (left) and quenched (right) randomness. As we previously stressed, in case of annealed randomness this dynamical feature has an important role on the final outcome and by changing only this parameter we can span from a full defection to a full cooperation state. Furthermore, this sensitivity on $w_l$ value remains valid independently of the applied $0<q<1$ value. But these features, as we argued above, disappear for the quenched mixing case. In the latter case only the value of $q$ counts. This is illustrated nicely on the right panel where we obtained higher cooperation level at {\it higher} temptation value for larger $q$.

\begin{figure}
\centering
\includegraphics[width=6.5cm]{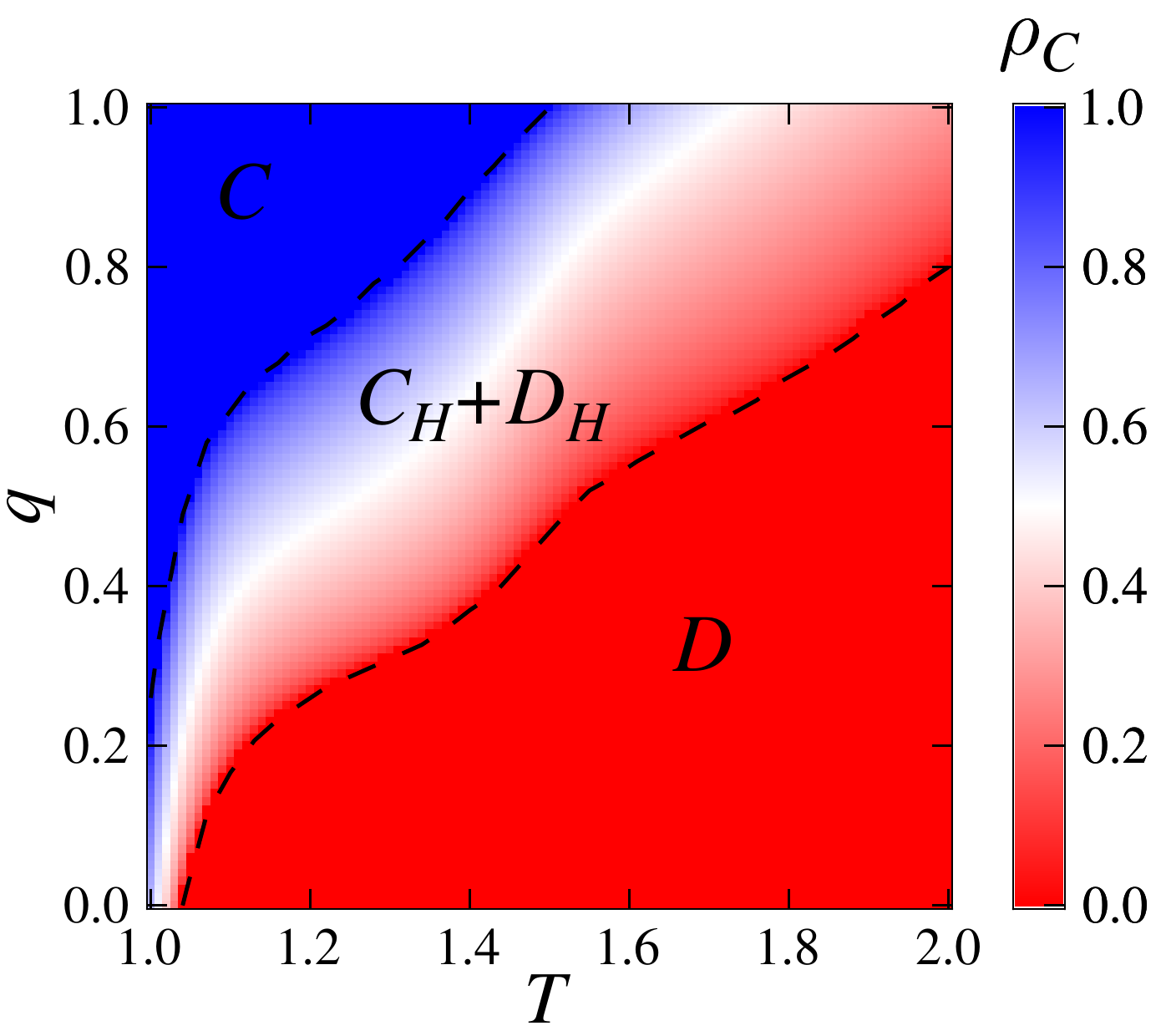}\includegraphics[width=6.5cm]{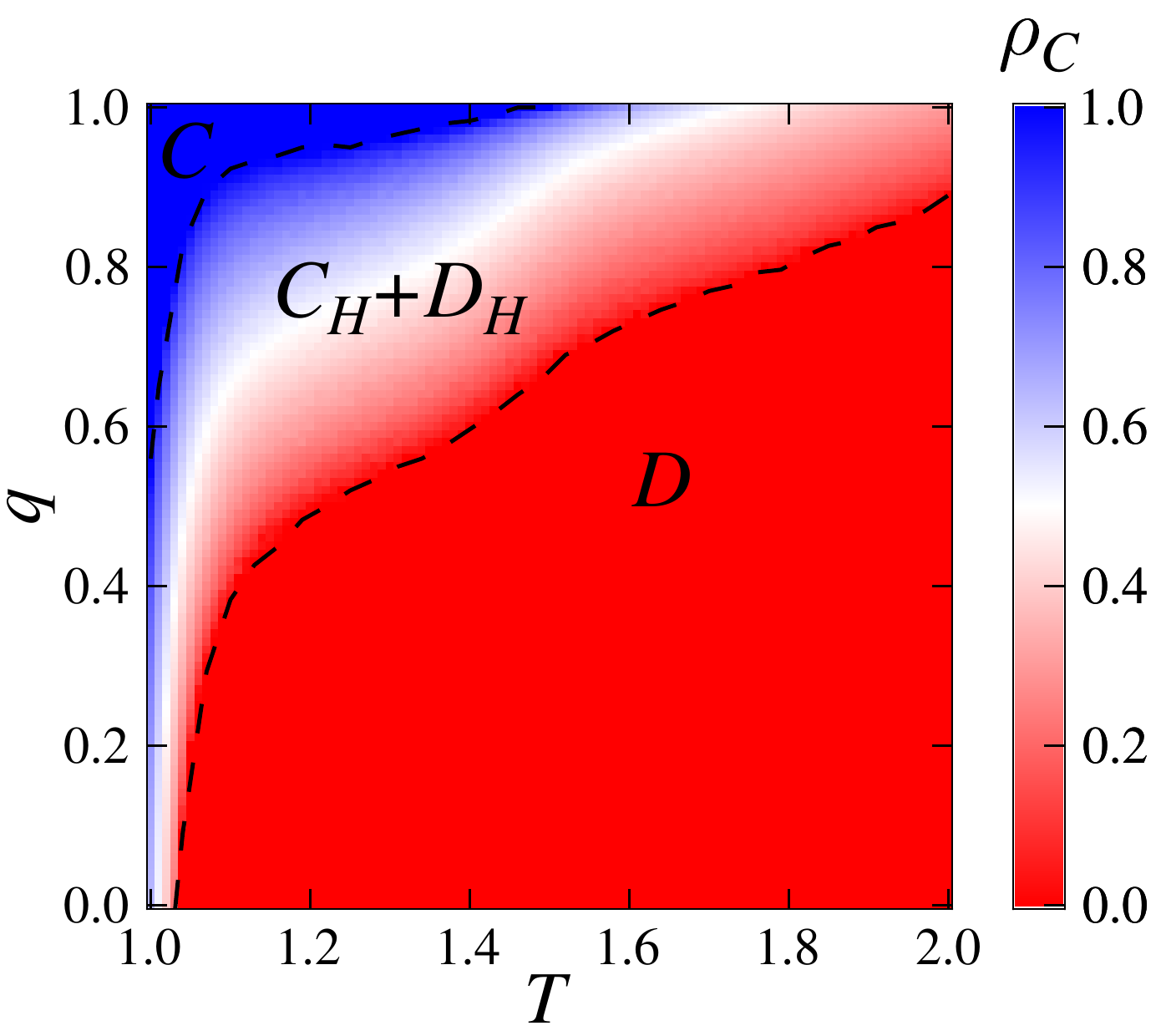}\\
\caption{Phase diagram of the coevolutionary model with heterogeneous teaching activity obtained for annealed (left) and quenched (right) mixture of updating rules when $w_t=0.1$ is applied beside the usual $w_t=1$ value. As for Fig.~\ref{phd_learn}, color codes illustrate the stationary cooperation level at specific parameter values of $q$ and $T$. In the intermediate phase, where cooperators and defectors coexist, only players with high teaching activity survive. The results were obtained up to $L=2000$ system size and remain robust in the large size limit.}\label{phd_teach}
\end{figure}

In the rest of this paper we discuss the cases when heterogeneous strategy teaching activity is assumed. This dynamical feature was proved to be relevant in previous studies \cite{szolnoki_epl07,szolnoki_epjb08,szolnoki_njp08}. Now, similarly to the learning cases, four different microscopic states compete for space during the coevolutionary process. They are cooperators with low ($C_L$) and high ($C_H$) teaching capacities and defectors with low ($D_L$) and high ($D_H$) convincing skill. The key observations are summarized in Fig.~\ref{phd_teach} for both cases. 

In case of annealed mixture of strategies, shown in the left panel, we can see that the previously reported spread of full cooperator state on the $q-T$ plane is limited. There are still full cooperator state in the high $q$ -- low $T$ corner, which is composed by $C_L$ and $C_H$ players, but the majority of parameter space is dominated by full defector phase. Between them, in the mixed phase, cooperators and defectors with high teaching ability coexist.

Based on our previous experience with learning activities, the case of quenched mixture, plotted on the right panel of Fig.\ref{phd_teach}, is less surprising. As previously, we can detect nonzero cooperation level only at high $q$ and low $T$ values. 

\begin{figure}
\centering
\includegraphics[width=13.5cm]{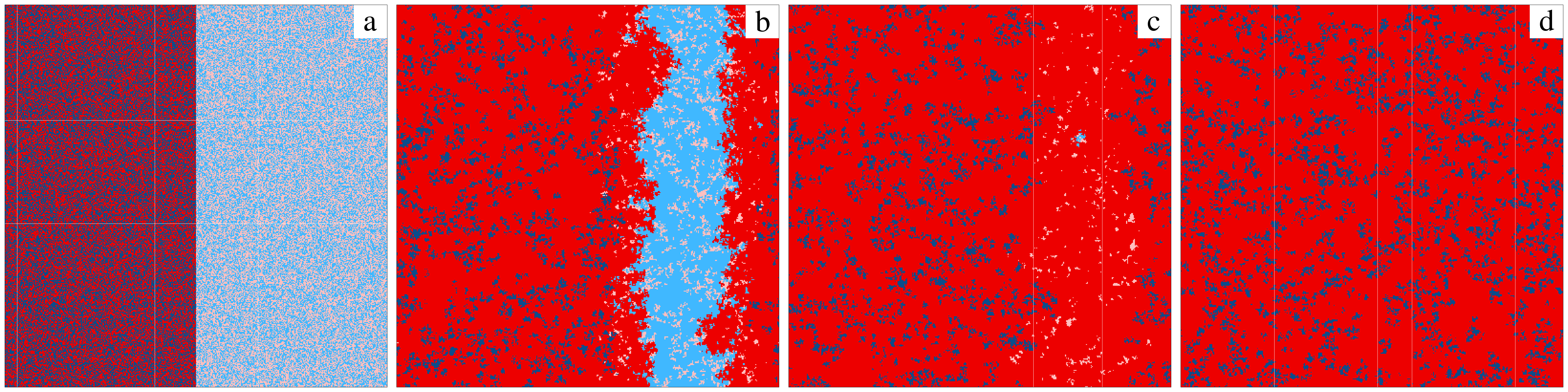}\\
\caption{Coevolution of strategies and individual teaching capacities at $T=1.4$ and $q=0.4$. Panel~(a) shows the starting state where players with high teaching capacities are distributed in the left side of the space. Here $D_H$ defector players (dark red) and $C_H$ cooperators (dark blue) are present. The right side of the initial panel is filled with players with low teaching capacities ($w_t=0.1$). They are $D_L$ defectors (light red) and $C_L$ cooperators (light blue). When evolution is launched, shown in panel~(b), two sub-solutions emerge where $D_H+C_H$ coexist on the left, while $D_L+C_L$ coexist on the right. The latter is not stable because $D_H$ players invade $C_L$ players very fast, leaving a full defector area which is composed by $D_H$ and $D_L$ players (shown in panel~(c)). However, this state is not stable again because $C_H$ players can invade the $D_L$ spots. The final state is composed by $D_H$ and $C_H$ players, but the fraction of the latter strategy is moderate. Linear system size is $L=400$.}\label{teach_a_t}
\end{figure}

To understand better the microscopic mechanisms which are responsible for these outcomes we present a series of snapshots obtained from an evolutionary process of annealed mixing in Fig.~\ref{teach_a_t}. Similar to the previous demonstration we use a prepared initial state again where players with high teaching activity are distributed randomly on the left side of panel~(a). They are $D_H$ (dark red) and $C_H$ (dark blue) players. On the right side of the starting panel players with low teaching capacities are present. They are $D_L$ (light blue) and $C_L$ (light red) players. Starting evolution from these random states we can observe that both pairs of cooperator states coexist with its own defector partner at this $T-q$ parameter values. Interestingly, the combination of $C_L+D_L$ provides a larger cooperation level than the combination of $C_H+D_H$ states. This is visible on panel~(b) where the former domain is almost blue while the latter is mostly red. This is the manifestation of the dynamic-sensitive cooperation we already reported when suppressed learning activities were used.

Rather unexpectedly, however, the solution with lower general cooperation level is more stable and gradually invades the other solution. Technically, it happens via the invasion of $D_H$ state which beats $C_L$ players who will die out first, as it is illustrated in panel~(c). The remaining $D_L$ spots are invaded by $C_H$ players and the system finally evolves into a state where $D_H$ and $C_H$ players coexist by giving a modest cooperation level. (The whole evolution can be followed in the animation we provided in \cite{teach}.) 

The above presented pattern formation explains why we obtain significantly less average cooperation in the left panel of Fig.~\ref{phd_teach} comparing to the left panel of Fig.~\ref{phd_learn}. Albeit a higher cooperation level would be available by using players with lower teaching ability but they are vulnerable against those who have higher teaching activity. This will result in the latter group's victory with a moderate cooperation level.

The comparison of the left panels of Fig.~\ref{phd_learn} and Fig.~\ref{phd_teach}(b) illustrates nicely our previous conclusion obtained at Fig.\ref{learn_compare}. More precisely, the mentioned phase diagrams for quenched randomness are practically identical, highlighting that the individual values of dynamical features have no relevant importance on the final outcome when updating rules are mixed in a quenched way. In the latter case the only decisive factor is the fraction of those sites where Death-Birth rule is applied: if this portion is above the percolation threshold then cooperators can survive at not too large $T$ values. Otherwise, when this portion is below this threshold then the system behavior is practically identical to the classic model when imitation is used by all players \cite{szabo_pre05}.

\section{Discussion}
\label{disc}
In this work we have explored the possible impact of multiple strategy updating rules on the cooperation level when players with different individual dynamical traits are present. The latter can be strategy learning or teaching capacities which determine the success of a microscopic imitation process. In a uniform system, where all players possess the same trait, these dynamical features have no relevant impact on the stationary state that is obtained as the final destination of an evolutionary process. Our work highlighted that this picture is inaccurate when different updating rules are present because the cooperation level may depend sensitively on the dynamical details. Interestingly, the individual learning skill of players are more important than the strategy teaching capacity and by varying the value of the former parameter we can reach a full cooperator state where payoff values would dictate a full defector state otherwise.

We also pointed out the way how to mix updating rules is also important. When a player can apply both imitation and Death-Birth rule via an annealed-like mixing then the above mentioned synergy can be observed. But this phenomenon is completely missing when the simultaneous presence of updating rules is realized in a way when different players use different rules permanently. The latter mix resembles a quenched randomness where we observe the sum of simpler subsystems using a singular updating rule. As a consequence, the system behavior is very similar to those we can see for traditional single-rule models. We stress that our observations about structured populations are robust and remain valid for non-regular interaction graphs as well.

It is worth noting that the simultaneous application of different updating rules within a single system is a recently opened direction \cite{danku_epl18,amaral_pre18} which may help to answer the long-standing debate whether which rule is evolutionary viable and which one captures the key element of a realistic system \cite{rand_pnas11,gracia-lazaro_srep12,rand_pnas14}. Our work warrants that several microscopic details can be important while other previously decisive conditions become irrelevant when multiple rules govern the evolution. We hope that our work will be a stimulating step for future research works along this path.

This research was supported by the Hungarian National Research Fund (Grant K-120785).

\bibliographystyle{elsarticle-num-names}

\end{document}